\documentclass[english,reprint,aps,amsmath,amssymb,prl,superscriptaddress,floatfix]{revtex4-2}
\usepackage{textcomp}

\usepackage[T1]{fontenc}
\usepackage{color}
\usepackage{bm}
\usepackage{varwidth}
\usepackage{amsmath}
\usepackage{amssymb}
\usepackage{graphicx}
\usepackage{soul}
\usepackage[normalem]{ulem}

\makeatletter

\usepackage{CJKutf8} 
\usepackage{siunitx}
\usepackage{epsfig}
\usepackage{multirow}
\usepackage{amsbsy}
\usepackage{array}
\usepackage{diagbox}
\usepackage{bm}
\usepackage{extarrows}
\usepackage{graphicx}
\usepackage{subfigure}
\usepackage{appendix}
\usepackage{txfonts}
\usepackage{lipsum}
\usepackage{bbding}
\usepackage{pifont}
\usepackage{makecell}
\usepackage{svg}
\usepackage{gensymb}
\usepackage{pifont}
\usepackage{dcolumn}

\usepackage{mathrsfs}
\usepackage[colorlinks,linkcolor=red,anchorcolor=,citecolor=green]{hyperref}
\allowdisplaybreaks[4]

\usepackage{amssymb}
\makeatletter

\newcommand{\Rmnum}[1]{\expandafter\@slowromancap\romannumeral #1@}
\makeatother

\makeatother

\usepackage{babel}
\newcommand{\m}[1]{\mathbf{#1}}

\begin{document}

\title{Direct probing the quantum geometric tensor for bosonic collective excitations}\

\author{Chi Wu}
\affiliation{Institute of Theoretical Physics, Chinese Academy of Sciences, Beijing 100190, China}
\affiliation{University of Chinese Academy of Sciences, Beijing 100049, China}

\author{Takashi Oka}
\affiliation{The Institute of Solid State Physics, The University of Tokyo, Kashiwa, Chiba 277-8581, Japan}

\author{Shuichi Murakami}
\affiliation{Department of Applied Physics, University of Tokyo,
7-3-1 Hongo, Bunkyo-ku, Tokyo 113-8656, Japan}
\affiliation{International Institute for Sustainability with Knotted Chiral Meta Matter (WPI-SKCM$^{2}$), Hiroshima University, Hiroshima, 739-8526, Japan}
\affiliation{Center for Emerged Matter Science, RIKEN, 2-1 Hirosawa, Wako, Saitama, 351-0198, Japan}

\author{Tiantian Zhang}
\email{ttzhang@itp.ac.cn}
\affiliation{Institute of Theoretical Physics, Chinese Academy of Sciences, Beijing 100190, China}

\begin{abstract}


The quantum geometric tensor (QGT), whose real and imaginary parts define the quantum metric and Berry curvature, encodes the intrinsic geometry of quantum states. 
While electronic QGT has recently become experimentally accessible and linked to diverse physical phenomena, its bosonic counterpart remains largely unexplored.
Here we show that the dynamical structure factor encodes the momentum-space structure of bosonic wave functions and thereby provides direct access to the full bosonic QGT throughout the Brillouin zone. Applying this framework, we uncover clear geometric signatures in the twofold quadrupole-Weyl phonon of BaPtGe and the nodal-line magnon in Gd, and further generalize the formalism to multiband systems. Our results establish a general route to measuring (non-)Abelian quantum geometry in bosonic systems, a crucial step toward elucidating its impact on condensed matter phenomena.


\end{abstract}

\maketitle

\textit{{Introduction.-{}-}}Quantum geometry tensor (QGT) characterizes the intrinsic Hilbert-space structure of quantum states; its real and imaginary parts define the quantum metric tensor (QMT) and Berry curvature, respectively~\cite{berry1984quantal,berry1989quantum,yu2025quantum,provost1980riemannian,cheng2010quantum,PhysRevA.108.032218}. 
Direct observation of the QGT and its components is essential for elucidating its impact on condensed matter phenomena. 
For example, the Berry curvature governs topological phases and transport responses\cite{thouless1982quantized,xiao2010berry,chang2008berry,sodemann2015quantum}, while the quantum metric in electronic systems has emerged as a key ingredient in phenomena ranging from nonlinear Hall effects~\cite{gao2023quantum,wang2023quantum,qiang2025clarification,gao2014field,han2024room,liu2021intrinsic} to flat-band superconductivity~\cite{xie2020topology,tian2023evidence,liang2017band,yu2024non,peotta2015superfluidity,julku2016geometric,herzog2022superfluid,hofmann2023superconductivity,peotta2023quantum,sun2024flat,kitamura2025quantumgeometrycorrelatedelectron}, anomalous Landau quantization~\cite{rhim2020quantum,hwang2021geometric,jung2024quantum,jung2024quantum}, and excitonic shifts~\cite{srivastava2015signatures,paiva2024shift,ying2024flat,verma2024geometric}, among others~\cite{neupert2013measuring,piechon2016geometric,gao2015geometrical,albert2016geometry,kolodrubetz2017geometry,ozawa2018steady,bleu2018effective,li2025unconventional,ouyang2025quantum,paul2024area,tam2024corner,zhou2024quantum,bzduvsek2025quantum,Pellitteri_2025,li2025quantum}. 
While the QMT has been experimentally accessed in artificial platforms~\cite{tan2019experimental,yu2020experimental,gianfrate2020measurement,sala2025quantum} and solids via techniques such as angle-resolved photoemission spectroscopy (ARPES), these measurements have so far been confined to electronic bands~\cite{kang2025measurements,kim2025direct}. Consequently, direct experimental access to the QMT, and thus the full QGT, for bosonic collective excitations remains elusive. Bridging this gap is a crucial step toward understanding the role of quantum geometry in bosonic systems~\cite{PhysRevLett.106.197202,li2024intrinsic,PhysRevB.108.165412,kuwabara2026formulationintrinsicnonlinearthermal}.

To overcome this challenge, we develop a general framework that enables direct access to the full QGT, including both the Berry curvature and quantum metric, of collective excitations via the dynamical structure factor (DSF)~\cite{sturm1993dynamic}. By formulating the DSF and QGT within a unified pseudospin framework, we establish a direct correspondence between measurable scattering intensities and the complete QGT. {The basic reconstruction is first illustrated in two-band subspaces and then extended to multiband pseudospin textures.} As a universal probe of density-fluctuation-driven quasiparticles, the DSF enables a practical protocol, implementable with inelastic x-ray or neutron scattering (IXS and INS), to resolve all components of the bosonic QGT across the Brillouin zone. {We demonstrate the approach using the twofold quadrupole-Weyl phonon in BaPtGe~\cite{zhang2020twofold,li2021observation} and the nodal-line magnon in Gd~\cite{scheie2022dirac}, and further extend it to spin-1 Weyl and charge-2 Dirac phonons~\cite{PhysRevLett.120.016401}, revealing clear signatures of bosonic quantum geometry.}

\begin{figure*}[!htbp]
\includegraphics[width=0.7\textwidth]{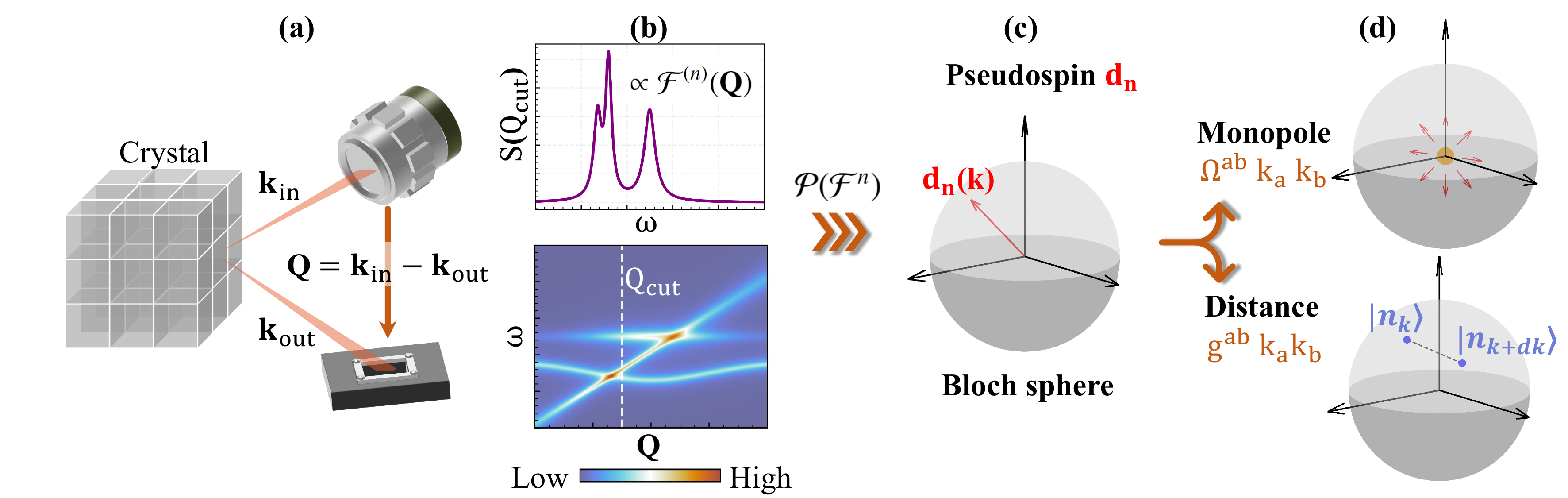}
\caption{\label{fig1} Schematic illustration of the direct measurement of the Berry curvature and quantum metric tensor via the dynamical structure factor $S(\mathbf{Q},\omega)$.
(a) Experimental geometry for measuring $S(\mathbf{Q},\omega)$.
(b) Measured dynamical structure factor $S(\mathbf{Q},\omega)$.
(c) Pseudospin texture reconstructed from $S(\mathbf{Q},\omega)$ or $\mathcal{F}$.
(d) Berry curvature and quantum metric tensor derived from the pseudospin texture. }
\end{figure*}

\textit{{Bosonic QGT.-{}-}}Generally, the Abelian QGT is defined in a $\mathrm{U(1)}$-gauge-invariant form as
\begin{equation}
\begin{aligned}
Q_{n}^{ab}(\mathbf{k}) &\equiv \langle \partial_a n_{\mathbf{k}}| (1 - \hat{P}_n) |\partial_b n_{\mathbf{k}} \rangle\\
&=\mathrm{Tr}\left[(\partial_a \hat P_n)(1-\hat P_n)(\partial_b \hat P_n)\right],
\end{aligned}\label{eq:qgt_1}
\end{equation}
where $\hat{P}_n=|n_{\mathbf{k}}\rangle\langle n_{\mathbf{k}}|$ is eigenprojector and $|n_{\mathbf{k}}\rangle$ is the periodic part of Bloch wavefunction in solids. Here $Q_{n}^{ab}$ denotes the band‑resolved (Abelian) QGT for a nondegenerate band, equivalently the U(1) QGT, while for a degenerate or strongly mixed $m$-dimensional subspace $\mathcal{M}$, one must promote $Q_{n}^{ab}$ to the U($m$)-gauge-invariant non‑Abelian QGT:
\begin{equation}
\begin{aligned}
Q_{\mathcal{M}}^{ab}(\mathbf{k}) &\equiv \mathrm{Tr}\left[\langle \partial_a u_{\mathbf{k}}| (1 - \hat{P}_\mathcal{M}) |\partial_b u_{\mathbf{k}} \rangle\right]\\
&=\mathrm{Tr}\left[(\partial_a \hat P_{\mathcal{M}})(1-\hat P_{\mathcal{M}})(\partial_b \hat P_{\mathcal{M}})\right],
\end{aligned}\label{eq:qgt_2}
\end{equation}
where $|u_{\mathbf{k}}\rangle=\oplus_{n\in\mathcal{M}}|n_{\mathbf{k}}\rangle$ and $\hat P_{\mathcal{M}}=\sum_{n\in\mathcal{M}}|n_{\mathbf{k}}\rangle\langle n_{\mathbf{k}}|$ equivalently characterize the subspace. 
Owing to the gauge invariance of eigenprojectors, the projector form provides a general QGT formulation, showing that quantum geometry is set by the parameter dependence of eigenprojectors. Below, QGT refers to the Abelian (band‑resolved) QGT used to characterize individual bands via their projectors; the non‑Abelian QGT for an $m$‑band subspace is defined analogously using the subspace projector.

For an $N$-band system $H_{\mathrm{eff}}(\mathbf{k})=h_0(\mathbf{k})\mathbb{I}_N+\mathbf{h}(\mathbf{k})\cdot\boldsymbol{\Lambda}$, where $\boldsymbol{\Lambda}$ denotes the $(N^2-1)$ generators of SU($N$), each eigenprojector corresponds to an SU($N$) pseudospin, $\hat P_n=\frac{1}{N}\mathbb{I}_N+\frac{1}{2}\mathbf{d}_n(\mathbf{k})\cdot\boldsymbol{\Lambda}$.
Hence the QGT can be written as~[\citenum{SM}]: 

{
\begin{equation}
Q_n^{ab}(\mathbf{k})= \underbrace{\frac{1}{4}\partial_a\mathbf{d}_n\cdot \partial_b\mathbf{d}_n}_{\displaystyle g^{ab}_n(\mathbf{k})}\,+\, \underbrace{\frac{i}{4}\mathbf{d}_n\cdot \left(\partial_a \mathbf{d}_n\times \partial_b \mathbf{d}_n\right)}_{\displaystyle -\frac{i}{2}\Omega_n^{ab}(\mathbf{k})}.
\label{eq:qgt_3}
\end{equation}
For a two‑band model $\mathbf{d}_{\pm}=\pm\mathbf{h}(\mathbf{k})/|\mathbf{h}(\mathbf{k})|$, hence knowledge of $\mathbf{d}_n$ fixes both QMT $g_n^{ab}$ and Berry curvature $\Omega_n^{ab}$. Unlike electronic systems, where circularly polarized ARPES can probe the pseudospin \cite{kang2025measurements,kim2025direct}, bosonic probes typically do not imprint polarization on the response, precluding an ARPES‑like strategy and necessitating a different approach.}
We propose the dynamical structure factor (DSF) as a direct probe of the bosonic QGT. While its explicit form depends on the specific system, the DSF admits a universal pseudospin representation. 

\textit{{Pseudospin Representation of the Phonon DSF.-{}-}}The DSF characterizes a system’s response to momentum- and energy-resolved probes, such as inelastic neutrons or x-rays (INS, IXS)~\cite{kotani2001resonant,ament2011resonant,benedek1994helium,hudson2006vibrational}. As illustrated in Figs.~\ref{fig1}(a)-(b), it encodes the spectrum of density fluctuations. For a phonon system, the total DSF for single-phonon coherent scattering reads
\begin{equation}
\begin{aligned}
S(\mathbf{Q},\omega)_{\text{1p}}=&\sum_{\mathbf{k},\sigma}\frac{N_{\text{cell}}}{\omega_{\mathbf{k},\sigma}}\Big|\sum_d\frac{f_d(\mathbf{Q})}{\sqrt{m_d}}e^{-W_d}\{\mathbf{Q}\cdot \mathbf{e}_{\sigma,d}(\mathbf{k})\}e^{i\mathbf{Q\cdot r}_d}\Big|^2\\
&\times\Big\{(\langle n_{\mathbf{k},\sigma}\rangle+1)\delta(\omega-\omega_{\sigma}(\mathbf{k}))\delta(\mathbf{Q-G-k})\\
&+\langle n_{\mathbf{k},\sigma}\rangle\delta(\omega+\omega_{\sigma}(\mathbf{k}))\delta(\mathbf{Q-G+k})\Big\}.
\end{aligned}
\end{equation}
Up to unimportant constant prefactors, the coherent contribution to the DSF for the $n^{th}$ phonon mode reads

\begin{equation}
\begin{aligned}
S_{\text{coh}}^{(n)}(\m{Q}, \omega) &\propto\mathcal{F}^{(n)}(\m{Q})\\
&= \left| \sum_{d}\frac{f_d(\mathbf{Q})}{\sqrt{m_d}}e^{-W_d} \m{Q} \cdot \mathbf{e}_{n,d}(\m{k}) e^{i \m{Q} \cdot \m{r_{d}}} \right|^{2}.
\end{aligned}\label{eq:DSF_1}
\end{equation}
Here, $\mathbf{Q}=\mathbf{G}+\mathbf{k}$ denotes the total momentum transfer, $f_d(\mathbf{Q})$, $m_d$, $W_d$, $\mathbf{r}_d$, and
$\mathbf{e}_{n,d}(\mathbf{k})$ denote the atomic form factor, atomic mass,
Debye-Waller factor, equilibrium position, and polarization vector of the
$d^{\text{th}}$ atom in the unit cell, respectively. For inelastic x-ray scattering, $f_d(\mathbf{Q})$ depends on momentum transfer $\mathbf{Q}$, whereas for neutron scattering it is a constant (the scattering length). 
Eq.~(\ref{eq:DSF_1}) reveals that the DSF is primarily governed by the factor $\mathbf{Q}\cdot\mathbf{e}_{n,d}(\m{k})$, which directly links the measured scattering spectra to the underlying microscopic dynamics, including the full phonon polarization structure.

To access the QGT through the DSF, one must first establish their connection via the pseudospin texture, as schematically shown in Fig.~\ref{fig1}. For simplicity, we assume that the target polarization vector involved in coherent scattering can be described by the eigenstate solution of the aforementioned $N$-band effective model, i.e., $H_{\text{eff}}|\psi_n\rangle=(h_0+\epsilon_n)|\psi_n\rangle$. Here, we adopt the following change of notation:
\begin{equation}
\begin{aligned}
\mathbf{e}_n&\to|\psi_n\rangle=\sum_s a_{n,s}|\xi_s\rangle,\\
\mathbf{e}_{n,d}&\to\sum_s a_{n,s}|\xi_{s,d}\rangle,\\
\end{aligned}
\end{equation}
where $\left\{|\xi_1\rangle,\ldots,|\xi_N\rangle\right\}$ are a set of orthonormal bases for the subspace.

{Given that the DSF scales with momentum transfer $\mathbf{Q}$ and experiments often probe $\mathbf{Q}$ beyond the first BZ, write $\mathbf{Q}\approx\mathbf{G}+\mathbf{C}$ with $\mathbf{G}$ a reciprocal lattice vector and $\mathbf{C}$ the BZ point (for $\Gamma$, $\mathbf{C}=0$). For measurements near $\Gamma$, one obtains}
\begin{equation}
\begin{aligned}
\mathcal{F}^{(n)}(\m{Q})&=\left|\sum_{s}\Bigg\{\Big\{ \sum_{d}\frac{f_d(\mathbf{Q})}{\sqrt{m_d}}e^{-W_d}e^{i \m{G} \cdot \m{r_{d}}}\m{G} \cdot |\xi_{s,d}\rangle\Big\}\cdot a_{n,s}(\mathbf{k})\Bigg\}\right|^{2}\\
&=\Big|\sum_s\langle W|\xi_s\rangle\langle\xi_s|\psi_n\rangle\Big|^2\\
&=V(\mathbf{G})\cdot\mathrm{Tr}[\hat P(W)\hat P_n]\\
&=V(\mathbf{G})\Big\{\frac{1}{N}+\frac{1}{2}\mathbf{n(G)}\cdot\mathbf{d}_n(\mathbf{k})\Big\},
\end{aligned}\label{eq:DSF_2}
\end{equation}
where $|W(\mathbf{G})\rangle$ is the measurement state encoding lattice and probe details. $\hat P(W)$ and $\hat P_n$ are the projectors onto $|W(\mathbf{G})\rangle$ and $|\psi_n\rangle$, respectively. $\mathbf{n}(\mathbf{G})$ is the pseudospin of the measurement state, defined as measurement axis, and $V(\mathbf{G})=\langle W|W\rangle$ is the squared norm of the measurement state. Thus, the intensity $\mathcal{F}^{(n)}(\mathbf{Q})$ equals the fidelity between measurement state and eigenstate and directly links to the pseudospin expressions for Berry curvature and the QMT. Eq.~\eqref{eq:DSF_2} is general for multiband systems; for magnons the same structure holds with a more compact measurement state due to fewer degrees of freedom.

 In practice, the reconstruction is governed by the measurement axes $\mathbf n(\mathbf{G})$. For a reduced momentum $\mathbf k$, measurements in different Brillouin zones, $\mathcal{F}^{(n)}(\mathbf G_i+\mathbf k)$, probe different projections of the same SU($N$) pseudospin $\mathbf d_n(\mathbf k)$ along $\mathbf n(\mathbf G_i)$. When the measurement matrix $M_{i\alpha}=n_\alpha(\mathbf{G}_i)$ has full column rank, the pseudospin can be reconstructed independently at each $\mathbf{k} $ point. If this condition is not fully satisfied (e.g., the experimental DSF data are sparse), the remaining degrees of freedom can be constrained either by intrinsic pseudospin/projector relations or through a global fit using symmetry-adapted basis functions~[\citenum{SM}].  The discrete pseudospin directly extracted from experimental DSF data already determines the QGT via finite-difference evaluation,  while symmetry-adapted fitting provides a smooth, symmetry-consistent reconstruction.

\begin{figure}[!htbp]
\centering
\includegraphics[width=0.48\textwidth]{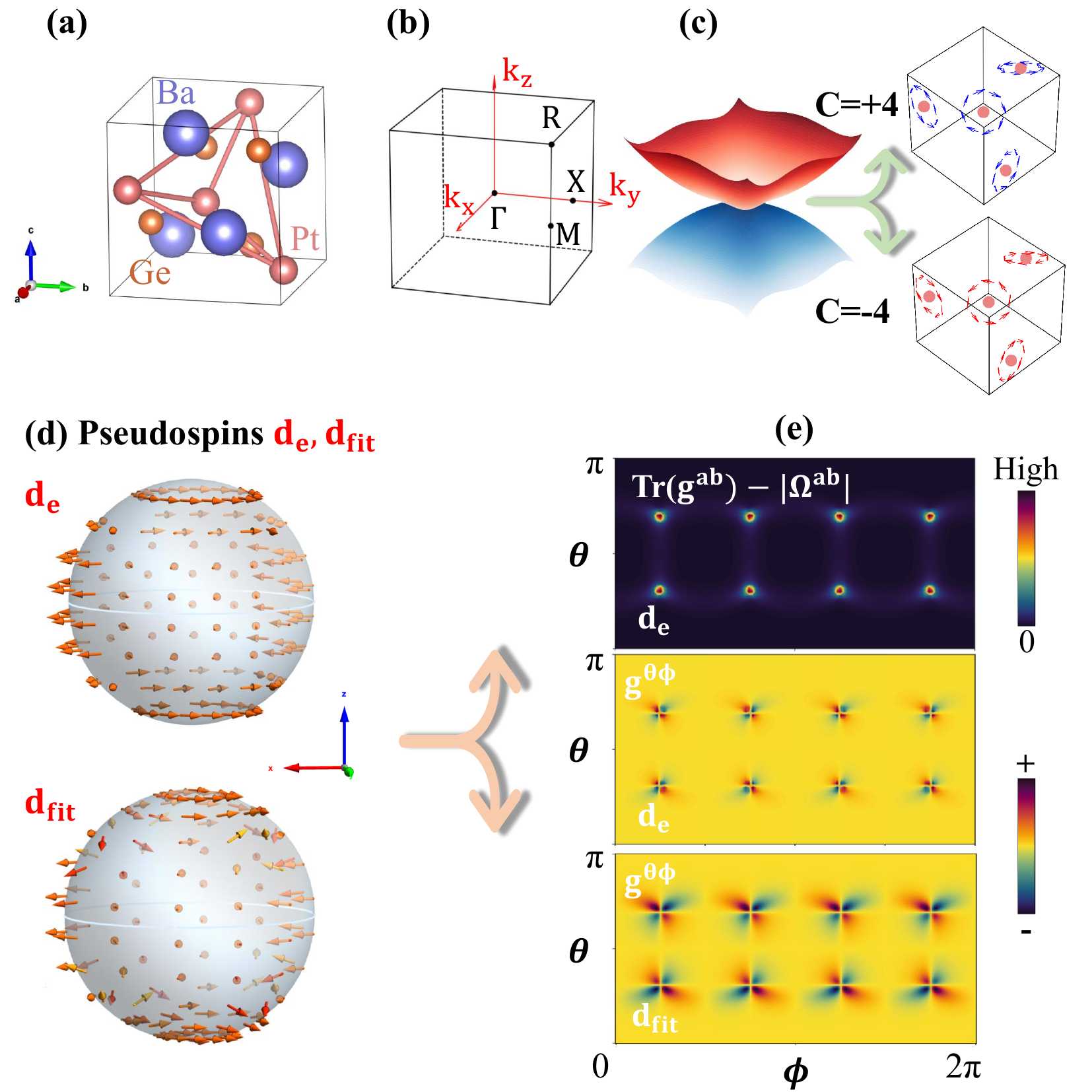}\caption{{(a) Crystal structure and (b) Brillouin zone for BaPtGe. (c) Phonon dispersion of the TQW and circularly polarized atomic motions for the two branches. (d) The pseudospin $\mathbf{d}_e$ directly extracted from DSF and the fitted pseudospin $\mathbf{d}_{\mathrm{fit}}$. (e) The positivity of $\left[\mathrm{Tr}(g^{ab})-|\Omega^{ab}|\right]$, evaluated from $\mathbf{d}_e$, verifies the topological bound; and the comparison of  QMT component obtained from $\mathbf{d}_e$ and $\mathbf{d}_{\mathrm{fit}}$, respectively.}}
    \label{fig:TWQ}
\end{figure}

\textit{{Phonon QGT in BaPtGe.-{}-}} As an example, consider the twofold quadrupole Weyl (TQW) phonon at $\Gamma$ in chiral cubic crystal BaPtGe~\cite{zhang2020twofold,li2021observation,jin2022chern} [Figs.~\ref{fig:TWQ}(a)-(c)]. Time‑reversal symmetry ($\mathcal{T}$) makes two 1D representations degenerate, yielding an isolated $N$=2 subspace. All atoms occupy the 4$a$ Wyckoff site, so we take basis functions
\begin{equation}
\begin{aligned}
|\xi_{1}\rangle&=|\xi_{1,\mathrm{Ba0}}\rangle\oplus\dots |\xi_{1,\mathrm{Pt0}}\rangle\oplus\dots |\xi_{1,\mathrm{Ge3}}\rangle,\text{and}\\
|\xi_2\rangle&=|\xi_1\rangle^*,
\end{aligned}
\end{equation}
where atoms sharing the same Wyckoff label have identical vibrations up to species-dependent complex amplitudes.  For example, for $\mathbf{r}_0=(c, c, c)$, $\frac{1}{A_{\mathrm{Ba}}}|\xi_{1,\mathrm{Ba0}}\rangle=\frac{1}{A_{\mathrm{Pt}}}|\xi_{1,\mathrm{Pt0}}\rangle=\frac{1}{A_{\mathrm{Ge}}}|\xi_{1,\mathrm{Ge0}}\rangle=\frac{1}{6}(1, \omega^2, \omega)$, and the Pt atom motions are shown in Fig.~\ref{fig:TWQ}(c). The amplitudes, determined by both atomic masses and interatomic coupling, can be obtained from DFT calculation.
In the INS, $f_d(\mathbf{Q})=b_{d,\text{coh}}$ (species‑dependent), so from Eq.~(\ref{eq:DSF_2}), the measurement components $\langle W|\xi_{1,2}\rangle$ and axis $\mathbf{n(G)}$ are fixed under this set of basis functions. For the TQW,
$\mathbf{d}\equiv\mathbf{d}_+=-\mathbf{d}_-$, the polarization $
\mathcal{P}(\mathbf{Q})\equiv\frac{\mathcal{F}^{(+)}-\mathcal{F}^{(-)}}{\mathcal{F}^{(+)}+\mathcal{F}^{(-)}}=\mathbf{n(Q)}\cdot\mathbf{d}(\mathbf{k})$ is directly obtained from measured intensities.We simulated INS via DFT‑convoluted $S^{\mathrm{r}}_{\mathrm{DFT}}(\mathbf{Q},\omega)$, integrated over $\Delta\omega=0.5$ meV for the two bands about $\mathbf{G}=(2,3,1), (2,5,0)$ and (5,3,0) on a sphere of radius 0.05 $\mathring{\mathrm{A}}^{-1}$. Fig.~\ref{fig:TWQ}(d) compares the pseudospin extracted directly from the DSF with that obtained from symmetry-adapted fitting, showing excellent agreement between the two.

For a 2D parameter manifold one has the bound $\mathrm{Tr}[g_{n}^{ab}]\geq 2\sqrt{\mathrm{det}(g_n^{ab})}\geq \left|\Omega_n^{ab}\right|$,
the second inequality saturating for two bands, and $\int [d\mathbf{k}]\sqrt{\text{det}(g_n^{ab})}=\pi C_n$~\cite{ozawa2021relations,peotta2015superfluidity,liu2025quantum,roy2014band}. The robustness of the experimental data can be assessed through the bound check [Fig.~\ref{fig:TWQ}(e)]. Besides, in 3D this lets one probe a node and its Chern number by evaluating the QMT on any closed 2D surface enclosing the node. The metric approach is especially useful for higher‑dimensional degeneracies: e.g. for a tensor monopole in 4D the 3‑form curvature is not the imaginary part of the conventional QGT, but the quantum volume from the QMT is directly related to the tensor‑monopole charge, so metric measurements provide a practical route to detect such topological charges~\cite{PhysRevLett.121.170401}.

\begin{figure}[!htbp]
\centering
\includegraphics[width=0.48\textwidth]{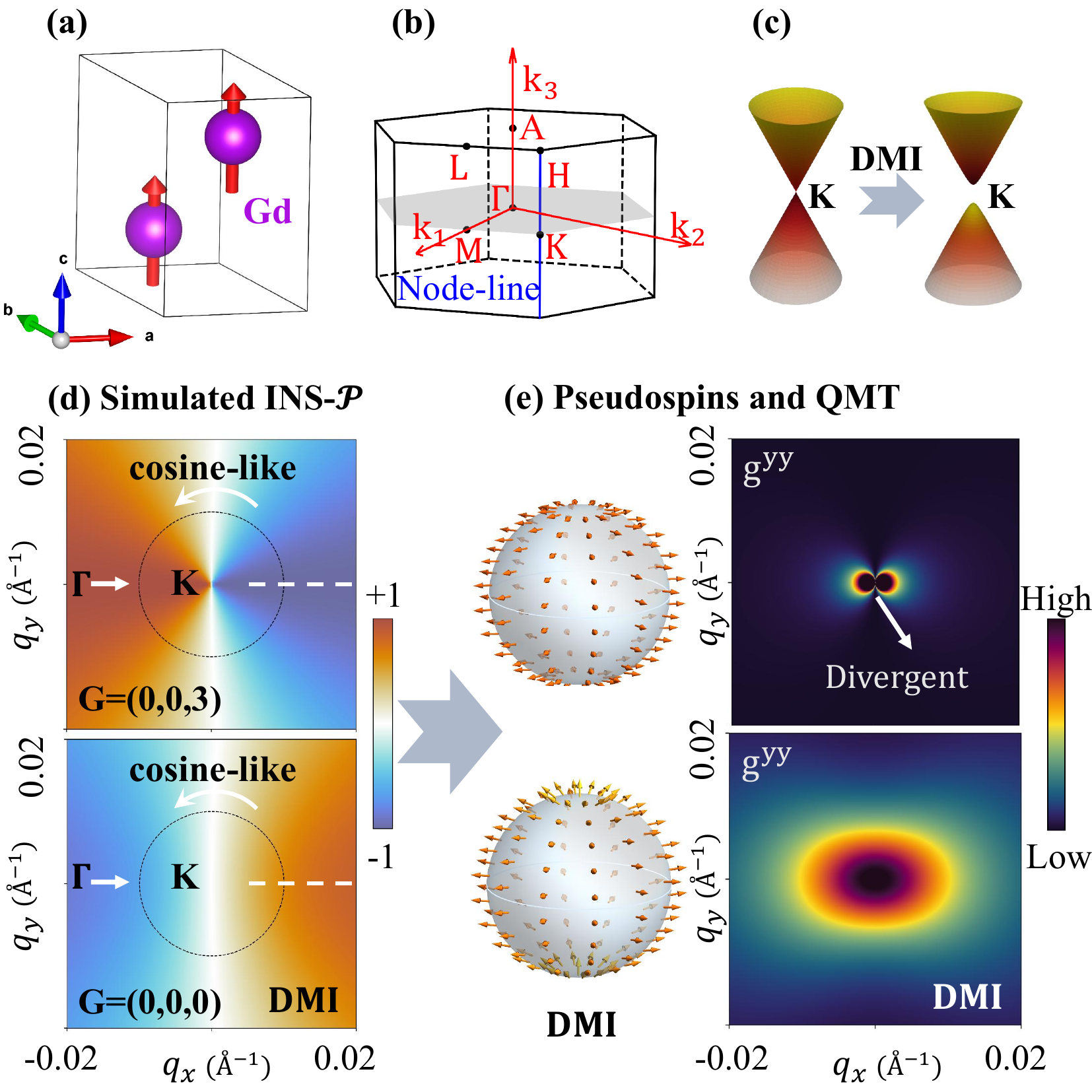}\caption{{ (a) Crystal structure and magnetic order of Gd. (b) Brillouin zone with the nodal‑line magnon. (c) Dirac cone at $K$ and DMI‑induced gap. (d) Heat maps of polarization $\mathcal{P}$ from energy‑integrated DSF on the $q_z$=0 plane: top without DMI [$\mathbf{G}=(0,0,3)$], bottom with DMI [$\mathbf{G}=(0,0,0)$], showing alternating intensity along $g_3$. (e) Fitted pseudospins and QMT components with/without DMI (full DSF used are in Ref.~[\citenum{SM}]). Without DMI, the fitted pseudospin $\mathbf{d}_{\mathrm{fit}}$ agrees well with the directly extracted pseudospin $\mathbf{d}_e$.}
}
\label{fig:magnon}
\end{figure}

\textit{{Magnon QGT in Gadolinium.-{}-}}Elemental gadolinium (Gd) has been reported to host both node-line magnons (along $H-K$) and node-plane magnons (on the {$q_z=\pi$} plane)~\cite{scheie2022dirac,scheie2022spin}, as shown in Figs.~\ref{fig:magnon}(a)-(b). Near the magnon Dirac cone at the $K$ point, INS experiments exhibit an anisotropic ``winding'' intensity pattern, which is captured qualitatively by linear spin-wave theory (LSWT)~\cite{scheie2022dirac}. This indicates that the two-band model constructed within the LSWT framework is reliable, allowing us to perform an analysis analogous to that for BaPtGe. Specifically, Bloch‑type basis functions at $K$ are constructed from the atomic magnon basis as:
\begin{equation}
\begin{aligned}
|\xi_1\rangle&=\frac{1}{\sqrt{N_{\text{cell}}}}\sum_{l}e^{i\mathbf{K\cdot R}_l}|l,\mathrm{Gd_1}\rangle,\\
|\xi_2\rangle&=\frac{1}{\sqrt{N_{\text{cell}}}}\sum_{l}e^{i\mathbf{K\cdot R}_l}|l,\mathrm{Gd_2}\rangle,\\
\end{aligned}\label{eq:basis_2}
\end{equation}
where $\mathbf{K}=\left(\frac{1}{3},\frac{1}{3},0\right)$ and $|l,\mathrm{Gd}_{d}\rangle=a^{\dagger}_{l,d}|\mathrm{FM}\rangle$ is the atomic magnon basis associated with the $d^{\text{th}}$ atom in the $l^{\text{th}}$ unit cell. Within this set of basis functions, the components of the measurement state take the compact form  $\langle \xi_s|W\rangle=e^{i\mathbf{Q\cdot r}_d}$, where $\mathbf{Q\approx G+K}$ and $\mathbf{G}=(g_1,g_2,g_3)$. The corresponding measurement axis is given by $\mathbf{n(Q)}=(-1)^{g_3}\left(\cos{\Delta_{12}},\sin{\Delta_{12}},0\right)$, where $\Delta_{12}=\frac{2\pi(g_2-g_1)}{3}$. From Eq.~(\ref{eq:DSF_2}) we can obtain the intensity and  polarization of the pseudospin along $\mathbf{n(Q)}$, which are given by
\begin{equation}
\begin{gathered}
S^{(\pm)}_{\text{coh}}(\mathbf{Q},\omega) \propto \mathcal{F}^{(\pm)}(\mathbf{Q}) = 1 \pm (-1)^{g_3}\left(d_x\cos{\Delta_{12}}+d_y\sin{\Delta_{12}}\right), \\
\mathcal{P}(\mathbf{Q}) = (-1)^{g_3}\left(d_x\cos{\Delta_{12}}+d_y\sin{\Delta_{12}}\right) ,
\end{gathered}\label{eq:DSF_3}
\end{equation}
where $\mathbf{q}$ is the effective wavevector measured from the $K$ point, and the factor $(-1)^{g_3}$ leads to alternating intensity between the upper and lower bands as the momentum crosses the Brillouin zone boundary along $g_3$ direction, offering a quantitative account of the observation in experiments~\cite{scheie2022dirac}.

No experimentally resolvable mode splitting in Gd bounds the total Dzyaloshinskii-Moriya exchange interaction (DMI) $\leq$3 $\mu$eV~\cite{scheie2022spin}. Thus, we simulated DSF for DMI=0 and 3 $\mu$eV and extracted polarization $\mathcal{P}(\mathbf{Q})$ from the energy‑integrated DSF [Fig.~\ref{fig:magnon}(d)]. 
Since Eq.~(\ref{eq:DSF_3}) gives only two pseudospin components, the measurement matrix $M_{i,\alpha}$ is not of full column rank and a low-order symmetry-adapted fitting is employed. Without DMI, the $D_{3h}$ little group at $K$ fixes $\mathbf{d(q)}=
(\cos\phi_{\mathbf{q}},\sin\phi_{\mathbf{q}},0)$, so fitting $\cos\phi_{\mathbf{q}}$ suffices, indicating that the directly extracted pseudospin data already contain the necessary information for the reconstruction;  $\phi_{\mathbf{q}}\approx\arg(q_x+iq_y)$ yields the observed cosine modulation [Fig.~\ref{fig:magnon}(e)].
With DMI, $C_{2x}$ is broken and $\mathbf{d}=(\cos\phi_{\mathbf{q}}\sin\theta_{\mathbf{q}},\sin\phi_{\mathbf{q}}\sin\theta_{\mathbf{q}},\cos\theta_{\mathbf{q}})$; the cosine modulation remains radius‑dependent due to $\theta_{\mathbf{q}}$, so we fit the unnormalized, symmetry‑constrained $\mathbf{d}$ [results in Fig.~\ref{fig:magnon}(e)].

We validated the reconstruction by computing topological charge from the fitted pseudospin. For the nodal‑line (DMI=0) the Berry phase reads $\gamma_{\pm}(C)=\mp \frac{1}{2} \oint_{C} \nabla_{\mathbf{q}} \phi_{\mathbf{q}} \cdot d\mathbf{q}
= \mp \pi\label{eq:Berry}$, in agreement with the Heisenberg–DM model. With 3 $\mu$eV DMI the Berry phase reconstructed from the fitted pseudospin yields the same phase for a sufficiently large loop $C$, confirming the Chern‑band around $K$. Thus the DSF cosine modulation directly encodes the Berry phase, providing a practical route to the QGT. Other perturbations, such as external magnetic fields, can be treated analogously by using the field-dependent LSWT eigenmodes in Eq.~(\ref{eq:DSF_2}), without changing the reconstruction principle.


\begin{figure}[!htbp]
\centering
\includegraphics[width=0.48\textwidth]{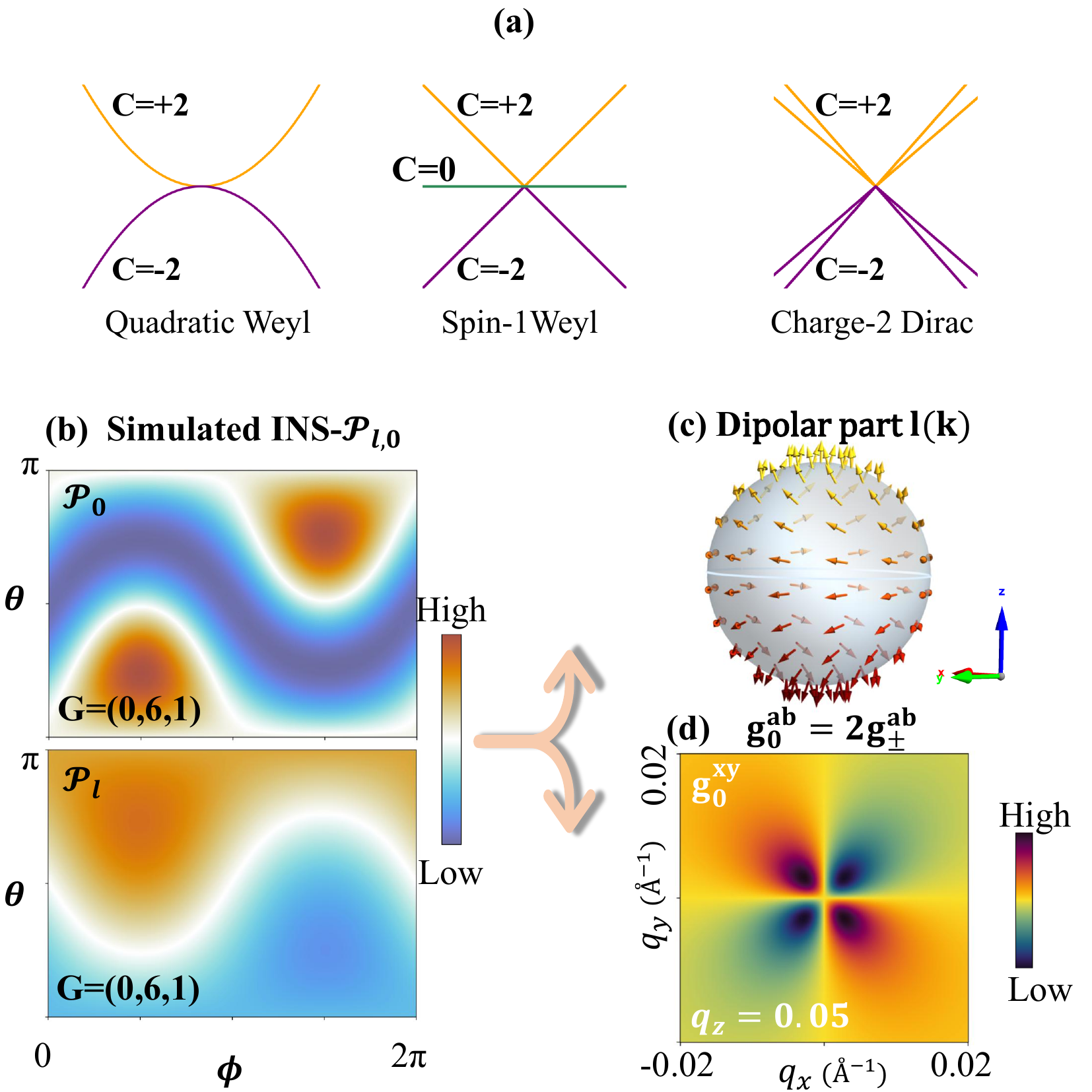}
\caption{{(a) Schematic illustrations of the three types of double Weyl points. (b) Heat maps of $\mathcal{P}_{l,0}$ from the energy-integrated DSF on a sphere of radius 0.01 $\mathring{\mathrm{A}}^{-1}$ about $\mathbf{G}$. (c) The pseudospin texture formed by the three nonzero components of the fitted dipolar part $\left(l_2(\mathbf{k}),l_5(\mathbf{k}),l_7(\mathbf{k})\right)$ and (d) corresponding QMT component for the spin-1 Weyl phonon. The full DSF data used are provided in Ref.~[\citenum{SM}]. }}
\label{fig:spin_1}
\end{figure}

\textit{Generalization to multiband systems.--}
For $N>2$, the central object remains the SU($N$) pseudospin, now with $N^2-1$ components subject to higher-dimensional projector constraints. Consequently, additional measurement axes or symmetry/projector constraints are generally required. Nevertheless, Eq.~(\ref{eq:DSF_2}) remains valid: the mode-resolved DSF measures the projection of the pseudospin onto the measurement axis. As concrete examples, we consider the spin-1 Weyl phonons at $\Gamma$ and the charge-2 Dirac phonons at $R$ in BaPtGe~\cite{PhysRevLett.120.016401}.

{ For the spin-1 Weyl phonon, let $\mathbf d_+$, $\mathbf d_0$, and $\mathbf d_-$ denote the SU(3) pseudospins of the upper, middle, and lower bands, respectively. The completeness relation of the projectors $\hat P_+ + \hat P_0 + \hat P_-=I_3$ implies $\mathbf d_+ + \mathbf d_0 + \mathbf d_-=0$. We therefore introduce the dipolar part $\mathbf l=(\mathbf d_+-\mathbf d_-)/2$, so that $\mathbf d_\pm=-\mathbf d_0/2\pm\mathbf l$. Eq.~(\ref{eq:DSF_2}) then shows that the three mode-resolved intensities directly determine
\begin{equation}
\begin{aligned}
\mathcal{P}_l(\mathbf{Q}) &\equiv \frac{\mathcal{F}^{(+)} - \mathcal{F}^{(-)}}{\mathcal{F}^{(+)} + \mathcal{F}^{(0)} + \mathcal{F}^{(-)}} = \mathbf{n(Q)} \cdot \mathbf{l}(\mathbf{k}),\\
\mathcal{P}_0(\mathbf{Q}) &\equiv \frac{2\mathcal{F}^{(0)} - \mathcal{F}^{(+)} - \mathcal{F}^{(-)}}{\mathcal{F}^{(+)} + \mathcal{F}^{(0)} + \mathcal{F}^{(-)}} = \frac{3}{2} \, \mathbf{n(Q)} \cdot \mathbf{d}_0(\mathbf{k}).
\end{aligned}
\end{equation}
Thus the DSF directly provides the projections needed to reconstruct the spin-1 projector texture. {By selecting eight distinct $\mathbf{G}$ that render the measurement matrix $M$ full column-rank, we directly extract the pseudospins $\mathbf{d}_0(\mathbf{k})$ and $\mathbf{l(k)}$. The fitted results show excellent agreement with the directly extracted pseudospins\cite{SM}.} Fig.~\ref{fig:spin_1}(b) shows the extracted $\mathcal P_l$ and $\mathcal P_0$ from the energy-integrated DFT-simulated DSF. The low-order fitted pseudospin texture gives the QMT and Chern numbers consistent with the spin-1 Weyl topology [Figs.~\ref{fig:spin_1}(c)].}

{For the charge-2 Dirac phonon, the four branches are not generically twofold degenerate around $R$; instead, symmetry protects twofold degeneracies only along certain high-symmetry paths, such as $X$--$R$--$M$. Along these lines, the projectors onto the two degenerate subspaces can be written as $P_{\pm}=\frac{1}{2}\mathbb{I}_4\pm\frac{1}{2}\mathbf{d(q)}\cdot\boldsymbol{\Gamma}$\cite{SM}, allowing the two-band analysis to be applied directly and non-Abelian QGT can be obtained~\cite{SM}. 
Away from these paths, the resolved branches carry well-defined Abelian QGTs, which can be directly obtained from the extracted pseudospins. Eq.~(\ref{eq:DSF_2}) remains valid for both cases. Moreover, since the non-Abelian QGT is well defined over the entire Brillouin zone, it is generally adopted as the relevant quantity for analysis.}


\textit{{Conclusions.-{}-}}In summary, we establish the dynamical structure factor as a direct experimental probe of both the Berry curvature and quantum metric, enabling complete access to the quantum geometric tensor across the Brillouin zone. Validated in multifold Weyl phonons and node-line magnons, our approach unifies experimentally measurable scattering intensities with quantum geometry through a pseudospin framework and is readily extendable to multi-band systems with well-defined pseudospin degrees of freedom. By directly linking the DSF to quantum geometry, our work enables experimental access not only to spectral properties but also to geometric and topological responses, including sensitivity to perturbations, topology-induced mode dynamics, enhanced transport, and Chern numbers, thereby establishing a general paradigm for exploring quantum-geometric phenomena in bosonic quasiparticles and paving the way for understanding its role in condensed matter systems.

\textit{{Note added.-{}-}}We note a related work on the resolution of topology and geometry from momentum-resolved spectroscopies \cite{huang2026resolution}.

\textit{{Acknowledgements.-{}-}}We acknowledge the helpful discussion with H. Miao.
T. Zhang and C. Wu acknowledge the support from National Key R\&D Project (Grant Nos. 2023YFA1407400 and 2024YFA1409200), the National Natural Science Foundation of China (Grant Nos. 12374165 and 12447101), and the Strategic Priority Research Program (B) of the Chinese Academy of Sciences (Grant Nos. XDB1720000 and XDB1270101). T.O. acknowledge support from JSPS KAKENHI (No. JP23H04865, No. 23K22487), MEXT, Japan. 
S.M. is supported by JSPS KAKENHI Grant Nos. JP22H00108 and JP24H02231, and also by  MEXT Initiative to Establish Next-generation Novel Integrated Circuits Centers (X-NICS) Grant No. JPJ011438.

\newpage
\nocite{*}

\bibliography{ref}

\end{document}